\documentclass[aps,twocolumn,preprintnumbers,amsmath,nofootinbib,amssymb]{revtex4}

\usepackage{graphics}
\usepackage{epsfig,epsf,psfrag}
\usepackage{times}
\usepackage{float}
\usepackage{color}
\usepackage{amsfonts,amssymb,stmaryrd,latexsym,amsmath}
\usepackage{multirow}
\usepackage{array}
\usepackage{hhline}
\usepackage{booktabs}
\usepackage{enumerate}
\usepackage{slashed}
\usepackage{subfigure}
\usepackage[colorlinks, citecolor=blue,anchorcolor=red,menucolor=red, linkcolor=red,filecolor=red,runcolor=red,urlcolor=blue,frenchlinks=red]{hyperref}
\usepackage{epstopdf}
\usepackage{braket}

\allowdisplaybreaks[4]

\begin{document}
	
\bibliographystyle{unsrt}

\title{Searching for doubly charmed tetraquark candidates $T_{cc}$ and $T_{cc\bar{s}}$ in $B_c$ decays}
	
\author{Yuan Li}
\author{Ying-Bo He}
\author{Xiao-Hai~Liu}~\email{xiaohai.liu@tju.edu.cn}
\author{Baoyi Chen}~\email{baoyi.chen@tju.edu.cn}
\author{Hong-Wei Ke}~\email{khw020056@tju.edu.cn}

\affiliation{ Center for Joint Quantum Studies and Department of Physics, School of Science, Tianjin University, Tianjin 300350, China	}
	
\date{\today}
	
\begin{abstract}
In this work, we propose to search for the exotic doubly charmed meson $T_{cc}^+$ and its analog $T_{cc\bar{s}}^+$ in $B_c^+$ decays, which provide a good environment for the formation of the exotic state containing double charm quarks. Within the molecular scheme, the production of $T_{cc}^+$ and $T_{cc\bar{s}}^+$ through various rescattering processes with different intermediate states are investigated. For the moderate values of model parameters, the branching ratios of $B_c^+$ decaying into $T_{cc}^+ \bar{D}^{0}$, $T_{cc}^+ \bar{D}^{*0}$, $T_{cc\bar{s}}^+ \bar{D}^{0}$ and $T_{cc\bar{s}}^+ \bar{D}^{*0}$ are estimated to be of the order of $10^{-7}$, $10^{-5}$, $10^{-6}$ and $10^{-4}$, respectively, which may be tested by future experiments. 		
\end{abstract}

\maketitle

\section{Introduction}
The LHCb collaboration recently reported the observation of a narrow doubly charmed tetraquark candidate $T_{cc}^+$ in the prompt production of proton-proton collisions~\cite{LHCb:2021vvq,LHCb:2021auc}. This is the first observation of the exotic state containing double charm quarks. The $T_{cc}^+$ is observed in the $D^0D^0\pi^+$ mass spectrum, and has mass of approximately 3875 MeV, which is just below the $D^{*+}D^0$ threshold. Using the Breit-Wigner parametrization, the location of the resonance peak relative to the $D^{*+}D^0$ threshold $\delta_m$ and the width $\Gamma$ are determined to be 
\begin{eqnarray}\label{Xstates}
		\delta_m &\equiv& m_{T_{cc}^+}-(m_{D^{*+}}+m_{D^0}) \nonumber \\
		&=& -273\pm 61\pm 5^{+11}_{-14}\ \mbox{keV}, \nonumber\\
	    \Gamma &=&	410\pm 165\pm 43^{+18}_{-38}\ \mbox{keV} ,
\end{eqnarray}
respectively. 

The long-lived $T_{cc}^+$ particle has the quark content $cc\bar{u}\bar{d}$ and the spin-parity quantum number $J^P=1^+$. The flavor quantum number is absolutely exotic. In fact, the tetraquark state with double heavy quarks has been studied for many years. There are two popular theoretical pictures describing the underlying structures of such tetraquark states, i.e., the compact tetraquark picture and hadronic molecule picture. In the compact tetraquark picture, the doubly heavy state is usually thought to be composed of the compact diquark and anti-diquark~\cite{Ballot:1983iv,Zouzou:1986qh,Carlson:1987hh,Heller:1986bt,Deng:2018kly,Yan:2018gik,Richard:2018yrm,Meng:2020knc,Meng:2021yjr,Qin:2020zlg,Navarra:2007yw,Dias:2011mi,Du:2012wp,Agaev:2019qqn,Feng:2013kea,Braaten:2020nwp,Cheng:2020wxa,Eichten:2017ffp,Guo:2021yws,Weng:2021hje,Kim:2022mpa,Agaev:2021vur,Karliner:2021wju}, while in the hadronic molecule picture, it is composed of a pair of heavy mesons~\cite{Manohar:1992nd,Tornqvist:1993ng,Ericson:1993wy,Janc:2004qn,Ding:2009vj,Molina:2010tx,Li:2012ss,Xu:2017tsr,Tang:2019nwv,Ohkoda:2012hv,Liu:2019stu,Ding:2020dio}. In Refs.~\cite{Tan:2020ldi,Vijande:2009kj,Yang:2009zzp,Yang:2019itm}, the compact tetraquark and hadronic molecule pictures are taken into account simultaneously. According to the LHCb measurements, the rather closeness of the $T_{cc}^+$ mass to the $D^{*+}D^0$ threshold strongly favors the molecular explanation concerning the $T_{cc}^+$ nature~\cite{Chen:2021vhg,Feijoo:2021ppq,Wang:2021yld,Ren:2021dsi,Deng:2021gnb,Albaladejo:2021vln,Dai:2021vgf,Du:2021zzh,Meng:2021jnw,Ling:2021bir,Yan:2021wdl,Fleming:2021wmk,Xin:2021wcr,Agaev:2022ast,Zhao:2021cvg,Ke:2021rxd}. This is similar to the case of the famous $X(3872)$ state, which is widely supposed to be a weakly bound hadronic molecule composed of $D^*\bar{D}$/$D\bar{D}^*$. We refer to Refs.~\cite{ParticleDataGroup:2022pth,Chen:2022asf,Guo:2017jvc,Kalashnikova:2018vkv,Brambilla:2019esw,Guo:2019twa,Chen:2016qju,Esposito:2016noz,Olsen:2017bmm,Lebed:2016hpi,Ali:2017jda,Liu:2019zoy} for a review concerning the recent progress of the doubly heavy tetraquarks and many other exotic hadrons discovered in the last two decades.

The production mechanism of hadrons is closely related to their intrinsic structures. In theoretical study, the production of $T_{cc}^+$ in the $\gamma p$ scattering, $pp$ collisions and heavy ion collisions have been discussed recently~\cite{Huang:2021urd,Braaten:2022elw,Hu:2021gdg,Abreu:2022lfy}. In 2008, Liu and Zhao have ever proposed an experimental scheme for searching for the doubly heavy mesons in the missing mass spectrum of two $\Lambda_c$ final states in nucleon-nucleon collisions~\cite{Liu:2008ck}. However, up to now the $T_{cc}^+$ is only observed in the prompt production of $pp$ collisions. Searching for the $T_{cc}^+$ in more reactions is important for both confirming its existence and understanding its nature. In this work, we propose to search for the $T_{cc}^+$ and its analog $T_{cc\bar{s}}^+$ in the $B_c$ decays.

This paper is organized as follows. After introduction, in Section II we present the formulae of the rescattering amplitudes corresponding to the $B_c^+\to T_{cc}^+\bar{D}^0$ and $B_c^+\to T_{cc}^+\bar{D}^{*0}$ processes. The pertinent numerical results and discussions are also given in Section II. In Section III, the similar processes $B_c^+\to T_{cc\bar{s}}^+\bar{D}^0$ and $B_c^+\to T_{cc\bar{s}}^+\bar{D}^{*0}$ are discussed. A summary is given in Section IV.

\section{Production of $T_{cc}^+$ in $B_c^+$ decays}\label{sec:II}
\begin{figure}[htb]
		\centering
		\includegraphics[width=0.85\linewidth]{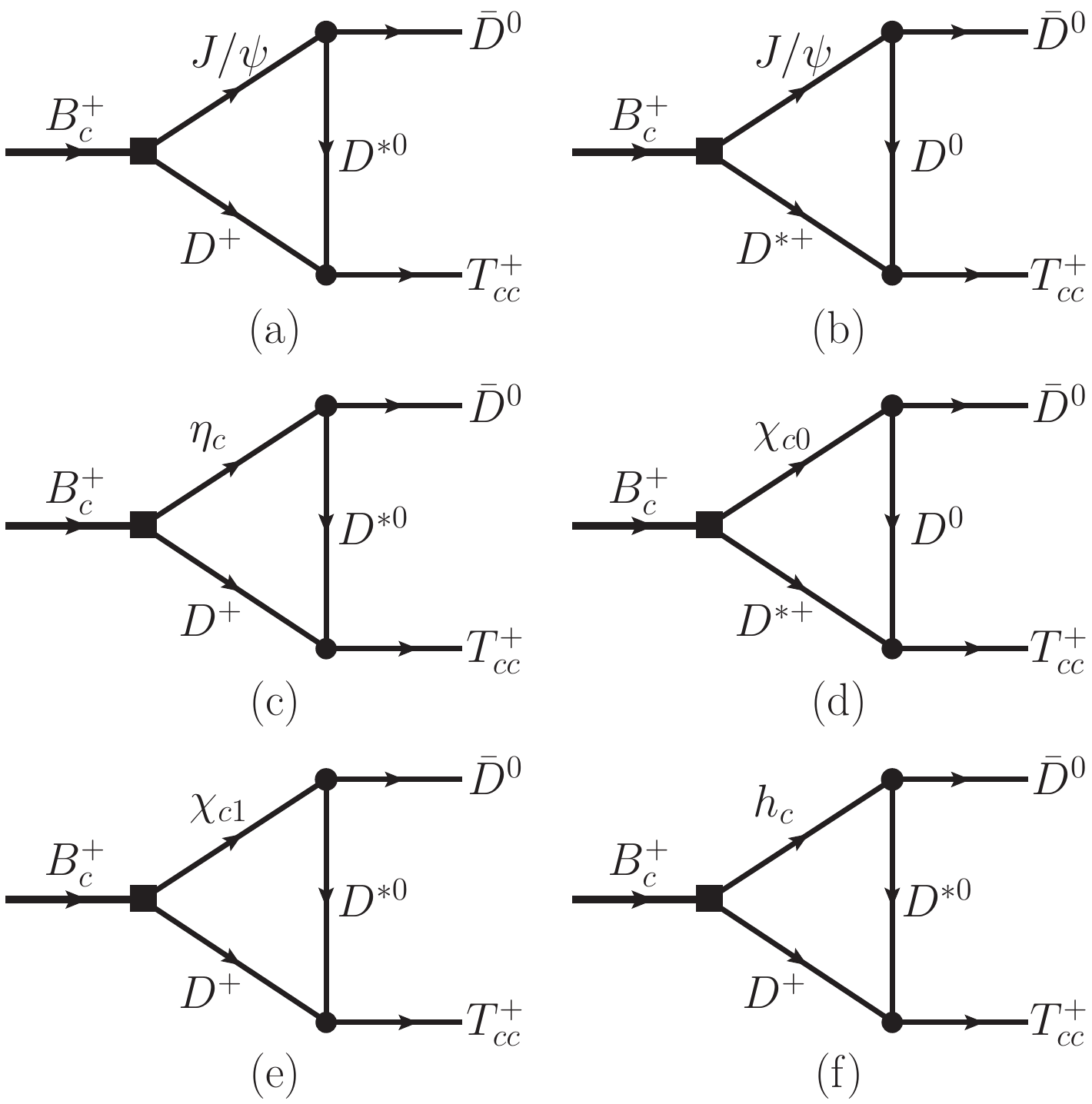}
		\caption{Rescattering diagrams contributing to $B_c^+\to T_{cc}^+\bar{D}^0$. The boxes and dots represent the weak and strong vertices, respectively.}
		\label{feyn:D0}
\end{figure}

\begin{figure}[htb]
		\centering
		\includegraphics[width=0.85\linewidth]{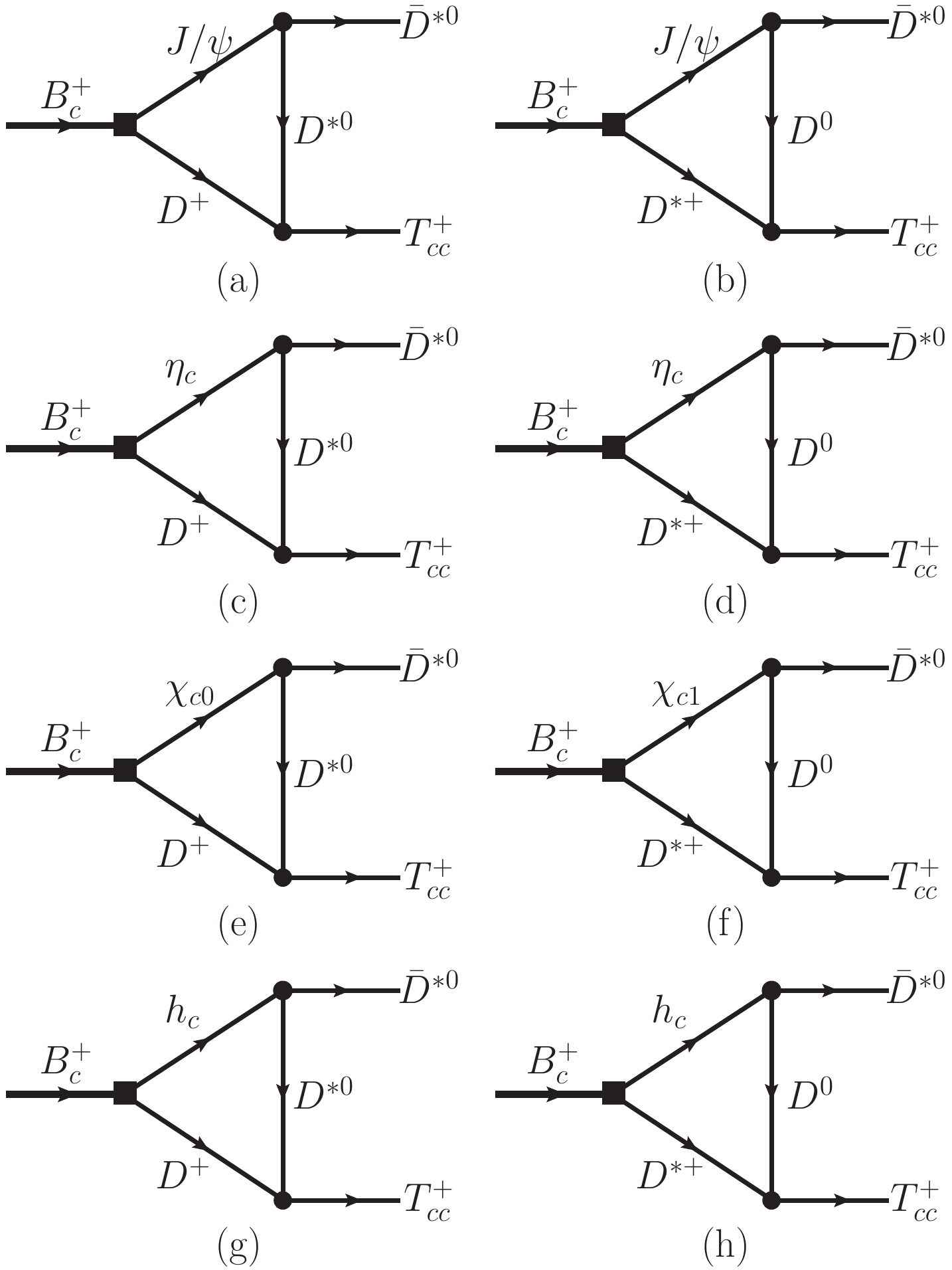}
		\caption{Rescattering diagrams contributing to $B_c^+\to T_{cc}^+\bar{D}^{*0}$. The boxes and dots represent the weak and strong vertices, respectively.}
		\label{feyn:Dstar0}
\end{figure}

If the $T_{cc}^+$ is a hadronic molecule, we can expect that it may be produced from the rescattering processes illustrated in Figs.~\ref{feyn:D0} and \ref{feyn:Dstar0}, i.e., the $B_c^+$ meson firstly decays into a charmonium and a charmed meson, the two particles then rescatter into the $\bar{D}^{(*)0}$ and the $T_{cc}^+$ via exchanging a charmed meson. In these rescattering diagrams, two charm quarks and one anti-charm quark are produced in the $B_c^+$ weak decay vertex, which creates a good environment for the formation of the double charm meson $T_{cc}^+$.

The $B_c$ decays also provide a good environment for the formation of the charmonium or charmonium-like states. In Refs.~\cite{Wu:2019vbk, Wang:2007sxa,Wang:2015rcz}, the production of tetraquark candidates $Z_c(3900)$, $Z_c(4020)$ and $X(3872)$ in $B_c$ decays has ever been discussed. The contributions from rescattering processes in weak decays are also widely discussed in the literature~\cite{Colangelo:2002mj,Colangelo:2003sa,Cheng:2004ru,Cheng:2005bg,Lu:2005mx,Wu:2021cyc,Liu:2020orv,Han:2021azw,Ge:2022dsp,Liu:2016onn,Liu:2017vsf}.


\subsection{Rescattering amplitudes}

At the quark level, the weak decay process $B_c^+\to M_{c\bar{c}} {D}^{(*)+} $ is induced by the $\bar{b}\to \bar{c} c\bar{d}$ transition, where $M_{c\bar{c}}$ represents a charmonium state. The charm quark in $B_c^+$ is treated as a spectator.
In the naive factorization approach the effective Hamiltonian governing the process reads
\begin{eqnarray}
H_{eff}&=&\frac{G_F}{\sqrt{2}}V_{cb}^* V_{cd}\big[c_1 (\bar{b}c)_{V-A}(\bar{c}d)_{V-A} \nonumber \\
&&+c_2 (\bar{c}c)_{V-A}(\bar{b}d)_{V-A} \big],
\end{eqnarray}
where $(\bar{q}_1 q_2)_{V-A}\equiv \bar{q}_1 \gamma_\mu (1-\gamma_5) q_2 $, $G_F$ is the Fermi constant, $V_{cb}$ and $V_{cd}$ are the CKM matrix elements, and $c_1$ and $c_2$ are the perturbatively calculable Wilson coefficients. Neglecting the contributions from the nonfactorizable, color-suppressed and annihilation terms, the decay amplitude of $B_c^+\to M_{c\bar{c}} {D}^{(*)+} $ can be factorized as 
\begin{eqnarray}\label{eq:fac}
&&\mathcal{A}(B_c^+\to M_{c\bar{c}} {D}^{(*)+})  \nonumber \\ &&=\frac{G_F}{\sqrt{2}}V_{cb}^* V_{cd} a_1 
\bra{M_{c\bar{c}}} \bar{b}\gamma_\mu (1-\gamma_5) c \ket{B_c^+} \nonumber \\
&&\times \bra{D^{(*)+}} \bar{c}\gamma^\mu (1-\gamma_5) d \ket{0},
\end{eqnarray}
where $a_1=c_1+c_2/N_c$, and $N_c$ is the number of colors. The factorized amplitude can be expressed in terms of the form factors of the transition $B_c^+\to M_{c\bar{c}}$ and the decay constant of $D^{(*)+}$.
The form factors of $B_c$ decaying into the lower charmonia have been well investigated in the literature~\cite{Wang:2008xt,Wang:2009mi,Zhang:2023ypl,Zhu:2017lqu,Wang:2018duy,Harrison:2020gvo,Nobes:2000pm,Ivanov:2005fd,Sun:2008ew,Dhir:2008hh,Kiselev:1999sc,Ivanov:2000aj,Zuo:2006re,Kiselev:2000pp,Du:1988ws,Ebert:2003cn,Hernandez:2006gt,Rui:2016opu,Ke:2013yka,ATLAS:2022aiy,LHCb:2013kwl,CDF:2012ksy,Colangelo:1999zn,Ivanov:2006ni,Dubnicka:2017job,Chang:1992pt}. To estimate the rescattering amplitudes, in this work we employ the relevant numerical results of Refs.~\cite{Wang:2008xt,Wang:2009mi}, where the form factors are calculated by means of the covariant light-front approach. The form factors of $B_c\to$ $J/\psi$ and $\eta_c$ induced by the vector and axial-vector currents are defined by
\begin{eqnarray}
&&\bra{\eta_c(k_1)} V_\mu \ket{B_c(p_0)}= \left( P_\mu -\frac{m_{B_c}^2-m_{\eta_c}^2}{q^2}q_\mu \right) \nonumber \\
&&\times F_1^{B_c \eta_c}(q^2)+ \frac{m_{B_c}^2-m_{\eta_c}^2}{q^2}q_\mu F_0^{B_c \eta_c}(q^2), \\
&&\bra{J/\psi(k_1,\varepsilon_{J}^*)} V_\mu \ket{B_c(p_0)}=-\frac{1}{m_{B_c}+m_{J/\psi}} \nonumber \\
&&\times \epsilon_{\mu\nu\alpha\beta}\varepsilon_{J}^{*\nu}P^\alpha q^\beta V^{B_c J/\psi}(q^2), \\
&&\bra{J/\psi(k_1,\varepsilon_{J}^*)} A_\mu \ket{B_c(p_0)}=i\bigg\{ (m_{B_c}+m_{J/\psi}) \varepsilon_{J}^{*\mu} \nonumber \\ &&\times A_1^{B_c J/\psi}(q^2) -\frac{P\cdot \varepsilon_J^{*}}{m_{B_c}+m_{J/\psi}} P_\mu A_2^{B_c J/\psi}(q^2) \nonumber \\
&&-2m_{J/\psi} \frac{P\cdot \varepsilon_J^{*}}{q^2} q_\mu [ A_3^{B_c J/\psi}(q^2)-A_0^{B_c J/\psi}(q^2)]\bigg\},
\end{eqnarray}
where $P=p_0+k_1$, $q=p_0-k_1$, and $\varepsilon_J$ is the polarization vector of $J/\psi$. The $B_c$ decaying to the $P$-wave charmonia $\chi_{c0}$, $\chi_{c1}$ and $h_c$ form factors are defined by
\begin{eqnarray}
&&\bra{\chi_{c0}(k_1)} A_\mu \ket{B_c(p_0)}= \left( P_\mu -\frac{m_{B_c}^2-m_{\chi_{c0}}^2}{q^2}q_\mu \right) \nonumber \\
&&\times F_1^{B_c \chi_{c0}}(q^2)+ \frac{m_{B_c}^2-m_{\chi_{c0}}^2}{q^2}q_\mu F_0^{B_c \chi_{c0}}(q^2), \\
&&\bra{A(k_1,\varepsilon_{A}^*)} A_\mu \ket{B_c(p_0)}=-\frac{1}{m_{B_c}-m_{A}} \nonumber \\
&&\times \epsilon_{\mu\nu\alpha\beta}\varepsilon_{A}^{*\nu}P^\alpha q^\beta A^{B_c A}(q^2), \\
&&\bra{A(k_1,\varepsilon_{A}^*)} V_\mu \ket{B_c(p_0)}=-i\bigg\{ (m_{B_c}-m_{A}) \varepsilon_{A}^{*\mu} \nonumber \\ &&\times V_1^{B_c A}(q^2) -\frac{P\cdot \varepsilon_J^{*}}{m_{B_c}-m_{A}} P_\mu V_2^{B_c A}(q^2) \nonumber \\
&&-2m_{A} \frac{P\cdot \varepsilon_A^{*}}{q^2} q_\mu [ V_3^{B_c A}(q^2)-V_0^{B_c A}(q^2)]\bigg\},
\end{eqnarray}
where the state $A$ represents the axial vector meson $\chi_{c1}$ or $h_c$.

The second matrix element in the right-hand side of Eq.~(\ref{eq:fac}) is parameterized as
\begin{eqnarray}
&&\bra{D^{+}(q)} \bar{c}\gamma^\mu (1-\gamma_5) d \ket{0}=i f_D q^{\mu}, \\
&&\bra{D^{*+}(q, \varepsilon_{D^*}^*)} \bar{c}\gamma^\mu (1-\gamma_5) d \ket{0}=f_{D^*}m_{D^*}\varepsilon^{*\mu}_{D^*},
\end{eqnarray}
with $f_D$ and $f_{D^*}$ the decay constants of $D$ and $D^*$, respectively.

One of the strong vertices in the rescattering diagrams of Figs.~\ref{feyn:D0} and \ref{feyn:Dstar0} involves a charmonium and a pair of open charm mesons. Taking into account the heavy quark spin symmetry, the effective Lagrangian describing these couplings is given by \cite{Colangelo:2003sa,Guo:2010ak,Casalbuoni:1996pg}
\begin{eqnarray}
&&\mathcal{L}_{\psi}=ig_\psi \braket{ 
\mathcal{J} \bar{H}_{2a}  \overleftrightarrow{\slashed{\partial}} \bar{H}_{1a}
} +\mbox{H.c.}, \label{lag:psi}\\
&& \mathcal{L}_{\chi}=ig_\chi \braket{ 
{\chi}^\mu \bar{H}_{2a}  \gamma_\mu \bar{H}_{1a}
} +\mbox{H.c.}, \label{lag:chi} 
\end{eqnarray}
with
\begin{eqnarray}
&& \mathcal{J} = \frac{1+{\rlap{v}/}}{2} [\psi^\mu \gamma_\mu-\eta_c\gamma_5]
\frac{1-{\rlap{v}/}}{2} , \\
&& \chi^\mu = \frac{1+{\rlap{v}/}}{2} \bigg[h_{c}^\mu \gamma_5+ \frac{1}{\sqrt{3}}(\gamma^\mu-v^\mu)\chi_{c0}  \nonumber \\
&& +\frac{1}{\sqrt{2}}\epsilon^{\mu\alpha\beta\gamma} v_\alpha \gamma_\beta \chi_{c1\gamma} +\chi_{c2}^{\mu\alpha}\gamma_\alpha 
\bigg]\frac{1-{\rlap{v}/}}{2} ,\\
&& H_{1a}  = \frac{1+{\rlap{v}/}}{2}[\mathcal{D}_{a\mu}^*\gamma^\mu-\mathcal{D}_a\gamma_5] , \\
&& H_{2a}  = [\bar{\mathcal{D}}_{a\mu}^*\gamma^\mu+\bar{\mathcal{D}}_a\gamma_5]\frac{1-{\rlap{v}/}}{2} , \\
&&\bar{H}_{1a,2a}=\gamma^0 H_{1a,2a}^{\dag} \gamma^0, \\ &&\mathcal{D}^{(*)}=(D^{(*)0},D^{(*)+},D^{(*)+}_s) ,
\end{eqnarray}
where $A\overleftrightarrow{\partial}_\mu B\equiv A({\partial}_\mu B) -(\partial_\mu A)B$, $H_{2a}$ is the charge conjugate field of $H_{1a}$, and $a$ is the light flavor index. The $S$- and $P$-wave charmonia are collected into the fields $\mathcal{J}$ and $\chi^\mu$, respectively. The pseudoscalar and vector charmed mesons are collected into the fields $H_{1}$ and $H_2$. All of the heavy fields in the above equations contain a factor $\sqrt{m}$ with $m$ the corresponding heavy meson mass. The $\braket{\cdots}$ in Eqs.~(\ref{lag:psi}) and (\ref{lag:chi}) means the trace over Dirac matrices. According to the effective Lagrangian $\mathcal{L}_\psi$ and $\mathcal{L}_\chi$, we can obtain the corresponding vertex function $\mathcal{A}(M_{c\bar{c}}\to \bar{D}^{(*)0} {D}^{(*)0})$ in the rescattering diagram. The relevant vertex functions are collected in Appendix \ref{sec:Appendix}. The coupling constants $g_\psi$ and $g_\chi$ can be estimated by invoking the vector meson dominance arguments~\cite{Colangelo:2002mj,Colangelo:2003sa}. The results are
\begin{eqnarray}
g_\psi &=& \frac{\sqrt{m_\psi}}{2m_D f_{\psi}}, \\
g_\chi &=& \sqrt{\frac{m_{\chi_{c0}}}{3}} \frac{1}{f_{\chi_{c0}}},
\end{eqnarray}
with $f_{\psi}$ and $f_{\chi_{c0}}$ the decay constants of $J/\psi$ and $\chi_{c0}$, respectively.

If we assume that the $T_{cc}^+$ is a pure isoscalar hadronic molecule, its wave function can be written as
\begin{eqnarray}
\ket{T_{cc}^+} = \frac{1}{\sqrt{2}}\left(\ket{D^{*+}D^0} -\ket{D^{*0}D^+}\right) .
\end{eqnarray}
The vertex functions in the rescattering diagrams for the $D^{*+}D^0$ or $D^{*0}D^+$ fusing into the $T_{cc}^+$ read
\begin{eqnarray}
\mathcal{A}(D^{*+}D^0\to T_{cc}^+)&=& +\frac{g_1}{\sqrt{2}} \varepsilon_{T_{cc}^+}^{*} \cdot \varepsilon_{D^{*+}}, \label{eq:g1}\\
\mathcal{A}(D^{*0}D^+\to T_{cc}^+)&=& -\frac{g_2}{\sqrt{2}} \varepsilon_{T_{cc}^+}^{*} \cdot \varepsilon_{D^{*0}}.\label{eq:g2}
\end{eqnarray}
Taking into account the isospin symmetry, the coupling constant $g_1$ should be equal to $g_2$. However, the thresholds of $D^{*+}D^0$ and $D^{*0}D^+$ are located above $m_{T_{cc}^+}$ around 0.3 and 1.7 MeV, respectively, and in the molecular scheme the coupling of the loosely bound state $T_{cc}^+$ with its component is relatively sensitive to the binding energy. Therefore we adopt the similar prescription used in Ref.~\cite{Meng:2021jnw} to account for the difference between $g_1$ and $g_2$. The coupling constants take the form
\begin{eqnarray}
g_{1,2}=\frac{4m_{T_{cc}^+}\sqrt{\pi \kappa_{1,2}}}{\sqrt{\mu_{1,2}}},
\end{eqnarray}
where $\kappa_1=\sqrt{2\mu_1(m_{D^{*+}}+m_{D^0}-m_{T_{cc}^+})}$, $\kappa_2=\sqrt{2\mu_2(m_{D^{*0}}+m_{D^+}-m_{T_{cc}^+})}$, and $\mu_1$ ($\mu_2$) is the reduced mass of $D^{*+}D^0$ ($D^{*0}D^+$).

The decay amplitude of $B_c^+\to T_{cc}^+\bar{D}^{(*)0}$ via one of the rescattering diagrams in Figs.~\ref{feyn:D0} and \ref{feyn:Dstar0} can be expressed in a general form as follows
\begin{eqnarray}\label{eq:loopamp}
&&\mathcal{A}(B_c^+\to T_{cc}^+\bar{D}^{(*)+})=-i\int \frac{d^4 k_3}{(2\pi)^4} \frac{\mathcal{A}(B_c^+\to M_{c\bar{c}} {D}^{(*)+})}{(k_1^2-m_1^2)} \nonumber \\
&&\times \frac{\mathcal{A}(M_{c\bar{c}}\to \bar{D}^{(*)0} {D}^{(*)0})\mathcal{A}(D^{*}D\to T_{cc}^+)}{(k_2^2-m_2^2)(k_3^2-m_3^2)} \mathcal{F}(k_3^2,m_3^2), 
\end{eqnarray}
where $k_{1}$, $k_{2}$ and $k_{3}$ ($m_{1}$, $m_{2}$ and $m_{3}$) correspond to the momenta (mass) of intermediate mesons $M_{c\bar{c}}$, $D^{(*)+}$ and ${D}^{(*)0}$, respectively.
The sum over polarization of intermediate state in Eq.~(\ref{eq:loopamp}) is implicit. For the intermediate spin-1 state, the sum over polarization reads
$\sum_{\mbox{pol}}\varepsilon_\mu \varepsilon_\nu^*=-g_{\mu\nu}+k_\mu k_\nu/m^2$. A dipole form factor $\mathcal{F}(k_3^2,m_3^2)$ is introduced in Eq.~(\ref{eq:loopamp}), which takes the form
\begin{eqnarray}\label{eq:formfactor}
\mathcal{F}(k_3^2,m_3^2)=\left(\frac{\Lambda^2-m_3^2}{\Lambda^2-k_3^2}\right)^2.
\end{eqnarray}
This form factor is supposed to parameterize the off-shell effects of the intermediate state and to kill the ultraviolet divergence in the loop integrals. We employ a dipole form factor rather than a monopole form factor because the latter cannot kill the divergence appearing in some rescattering amplitudes. The loop integral is performed by employing the program package \textit{LoopTools}~\cite{Hahn:1998yk}.

\subsection{Numerical analysis}
In this subsection, we give the results from explicit calculations of the rescattering amplitudes. First we list the input parameters used in the numerical calculation. In the weak decay amplitude $\mathcal{A}(B_c^+\to M_{c\bar{c}} {D}^{(*)+}) $, the central values of the CKM matrix elements $|V_{cb}|=0.0408$ and $|V_{cd}|=0.221$ reported by the Particle Data Group (PDG)~\cite{ParticleDataGroup:2022pth} are adopted, and the combination of Wilson coefficients $a_1$ is set to be 1.14~\cite{Ivanov:2006ni}. For the relevant decay constants, we use the following values: $f_D=0.209$ GeV~\cite{ParticleDataGroup:2022pth}, $f_{D^*}=0.245$ GeV~\cite{Becirevic:1998ua}, $f_\psi=0.416$ GeV~\cite{ParticleDataGroup:2022pth} and $f_{\chi_{c0}}=0.510$ GeV~\cite{Colangelo:2002mj}. Concerning the relevant particle mass, the PDG 2022 central values are used~\cite{ParticleDataGroup:2022pth}.

The numerical results of $B_c\to$ $\eta_c$, $J/\psi$, $\chi_{c0,c1}$ and $h_c$ form factors in Table II of Ref.~\cite{Wang:2008xt} and Table I of Ref.~\cite{Wang:2009mi} are employed. In our calculation of the loop integral, as an approximation we do not take into account the $q^2$-dependence of the form factors. The values of these weak decay form factors are fixed at $q^2=m_{D^{(*)+}}^2$, i.e., an on-shell approximation of the weak vertex function is adopted. 

\begin{figure}[tb]
		\centering
		\includegraphics[width=0.8\linewidth]{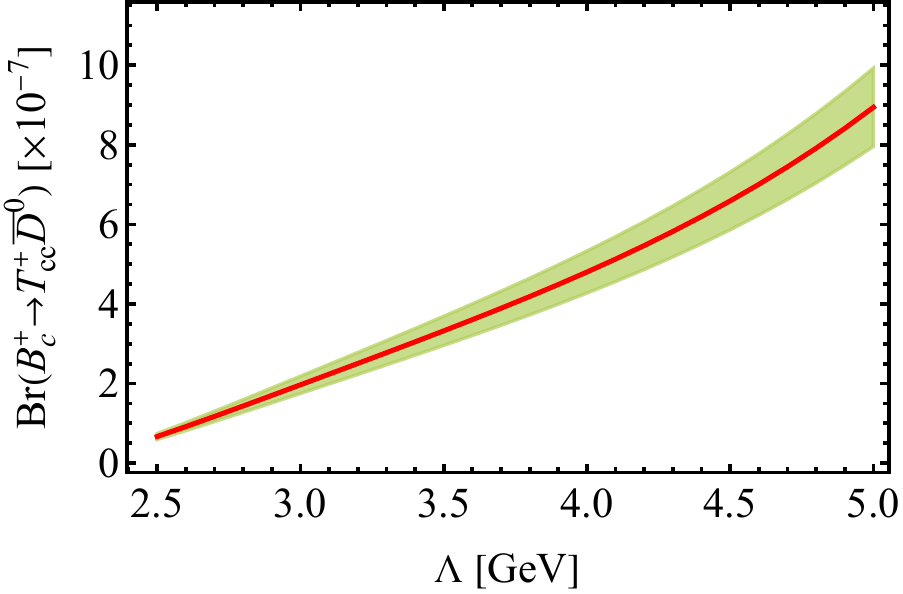}
		\caption{$\Lambda$-dependence of the branching ratio of $B_c^+\to T_{cc}^+ \bar{D}^0 $ via the rescattering processes in Fig.~\ref{feyn:D0}. The band is obtained by taking into account the uncertainties of the $\delta_m$.}
		\label{fig:D0Tcc}
\end{figure}

\begin{figure}[tb]
		\centering
		\includegraphics[width=0.8\linewidth]{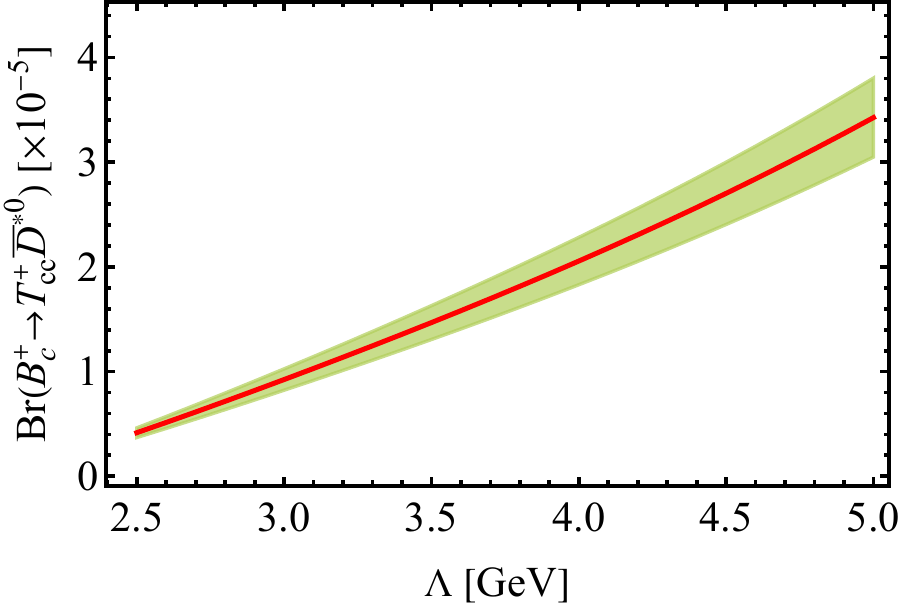}
		\caption{$\Lambda$-dependence of the branching ratio of $B_c^+\to T_{cc}^+ \bar{D}^{*0} $ via the rescattering processes in Fig.~\ref{feyn:Dstar0}. The band is obtained by taking into account the uncertainties of the $\delta_m$.}
		\label{fig:Dstar0Tcc}
\end{figure}

There is still a free parameter, i.e., the cutoff energy $\Lambda$, in the dipole form factor Eq.~(\ref{eq:formfactor}). Its explicit value should be determined from the experimental data. As an theoretical prediction, the empirical value of $\Lambda$ is usually set to be larger than the mass of the exchanged particle, and it also depends on the formalism of the form factor. For instance, in Ref.~\cite{Colangelo:2002mj}, one obtains $\Lambda\approx 2.7$ GeV to roughly fit the experimental data of $\mbox{Br}(B^-\to K^-\chi_{c0})$, where the form factor is the monopole type and the exchanged particles are $D$ and $D^*$. This cut-off energy is just around the typical values of the mass of the radially excited states of $D^{(*)}$ mesons. Taking into account the uncertainty induced by the cutoff $\Lambda$, we plot the branching ratio of $B_c^+\to T_{cc}^+ \bar{D}^{(*)0} $ via the rescattering processes as a function of $\Lambda$. The numerical results are displayed in Figs.~\ref{fig:D0Tcc} and \ref{fig:Dstar0Tcc}, where we consider a relatively larger range $2.5$--$5$ GeV for $\Lambda$. The uncertainties from the $T_{cc}^+$ mass are also taken into account.

The thresholds of final states $T_{cc}^+ \bar{D}^{0}$ and $T_{cc}^+ \bar{D}^{*0}$ are around $5739.7$ and $5881.7$ MeV, respectively, which are not far from the $B_c^+$ mass $6274.5$ MeV. Therefore the higher partial-wave amplitudes of the two channels are expected to be highly suppressed by the limited phase space.

From Figs.~\ref{fig:D0Tcc} and \ref{fig:Dstar0Tcc}, we can see that the two branching ratios increase monotonically with $\Lambda$ increasing. Within the cutoff range $2.5$--$5$ GeV, the branching ratio of $B_c^+\to T_{cc}^+ \bar{D}^0 $ is of the order of $10^{-7}$, while that of $B_c^+\to T_{cc}^+ \bar{D}^{*0} $ increases from $\mathcal{O}(10^{-6})$ to $\mathcal{O}(10^{-5})$ with $\Lambda$ increasing. The branching ratio of $B_c^+\to T_{cc}^+ \bar{D}^{*0} $ is much larger than that of $B_c^+\to T_{cc}^+ \bar{D}^0 $. This is because that the $S$-wave decay is allowed in $B_c^+\to T_{cc}^+ \bar{D}^{*0} $ but forbidden in $B_c^+\to T_{cc}^+ \bar{D}^0 $ to conserve the angular momentum. For the moderate cutoff $\Lambda$ around 3 GeV, $\mbox{Br}(B_c^+\to T_{cc}^+ \bar{D}^{*0})$ is of the order of $10^{-5}$, and we can expect it may be detectable in future experiments.

\section{Production of $T_{cc\bar{s}}^+$ in $B_c^+$ decays}
    \begin{figure}[htb]
		\centering
		\includegraphics[width=0.85\linewidth]{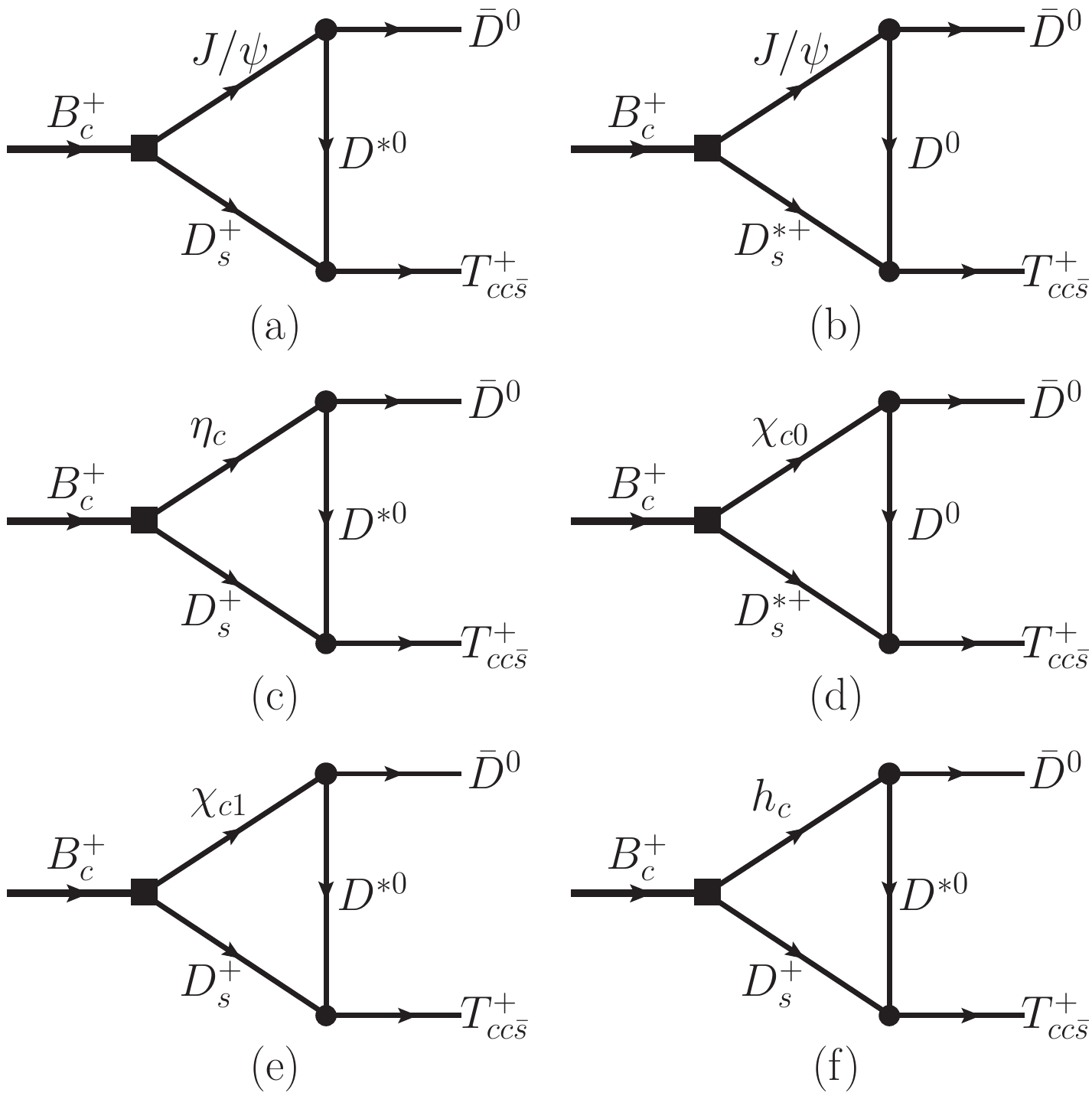}
		\caption{Rescattering diagrams contributing to $B_c^+\to T_{cc\bar{s}}^+\bar{D}^0$.}
		\label{feyn:TccsD}
	\end{figure}
	
	\begin{figure}[htb]
		\centering
		\includegraphics[width=0.85\linewidth]{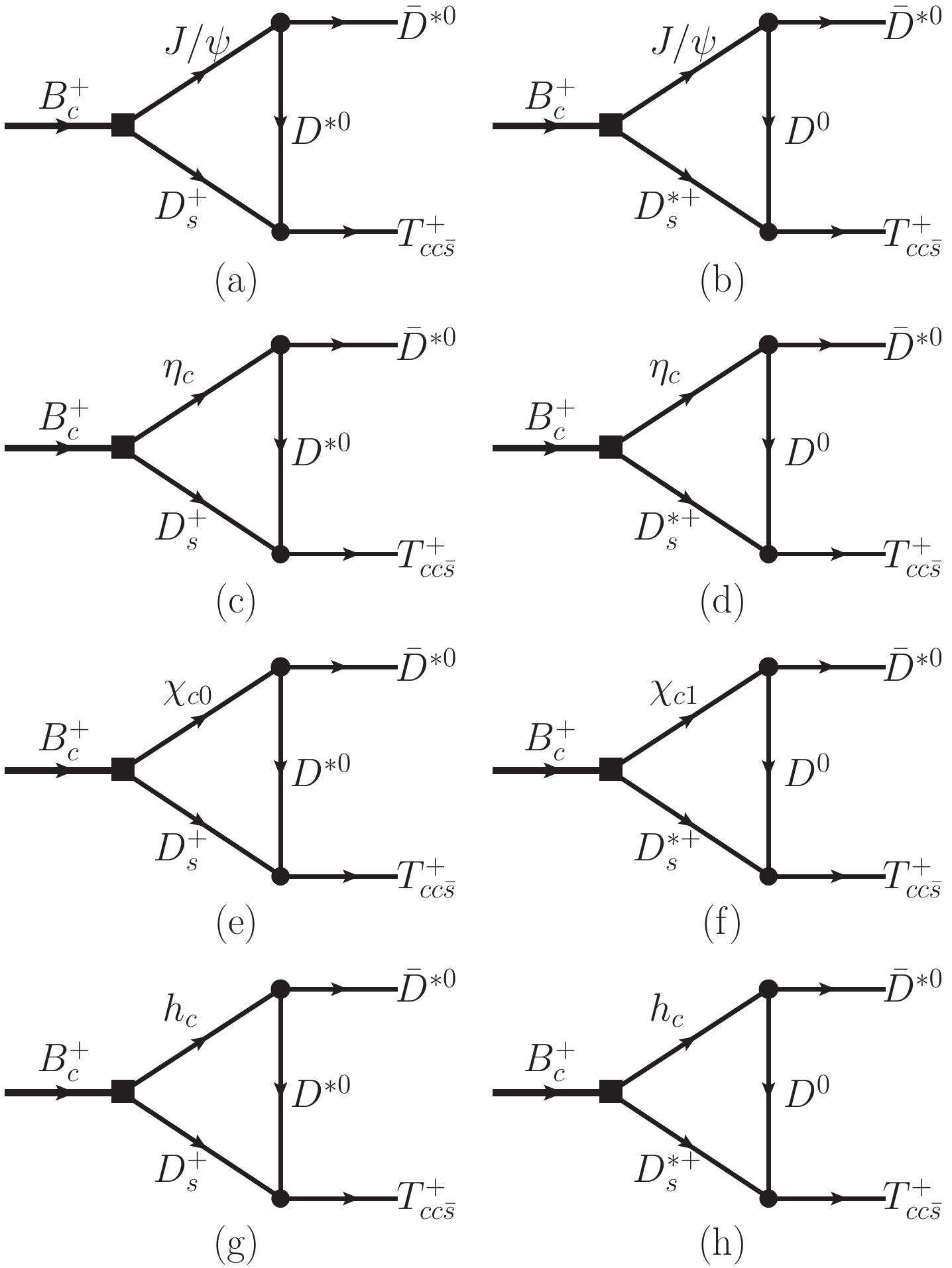}
		\caption{Rescattering diagrams contributing to $B_c^+\to T_{cc\bar{s}}^+\bar{D}^{*0}$.}
		\label{feyn:TccsDstar}
	\end{figure}

Inspired by the observation of $T_{cc}^+$, one may guess some other analogs can also exist, such as the doubly heavy states with the strange quark. In this work, we are interested in a $T_{cc}^+$ analog named $T_{cc\bar{s}}^+$. We assume $T_{cc\bar{s}}^+$ is a hadronic molecule composed of $D^{*+}_s D^0$/$D^+_s D^{*0}$, and the wave function is defined as
\begin{eqnarray}
\ket{T_{cc\bar{s}}^+} = \frac{1}{\sqrt{2}}\left(\ket{D^{*+}_s D^0} -\ket{ D^{*0} D^+_s}\right).
\end{eqnarray}
The $T_{cc\bar{s}}^+$ mass relative to the lower $D^{*0}D_s^+$ threshold $\delta_m$ is defined as
\begin{eqnarray}\label{eq:TccWF}
		\delta_m &\equiv& m_{T_{cc\bar{s}}^+}-(m_{D^{*0}}+m_{D^{+}_s}).
\end{eqnarray}
The relation between $T_{cc}^+$ and $T_{cc\bar{s}}^+$ is similar to that between $Z_c(3900)$ and $Z_{cs}(3985)$, which are widely supposed to be the hadronic molecules composed of $D\bar{D}^*/D^*\bar{D}$ and $D_s\bar{D}^*/D_s^*\bar{D}$.
In Ref.~\cite{Ding:2020dio}, the doubly heavy systems composed of a pair of heavy mesons have been systematically studied in a quasi-potential Bethe-Salpeter equation approach. The authors predict that the bound state with $J^P=1^+$ can be found from the $D_s^*D$-$D_s D^*$ coupled channel interactions, i.e., the hadronic molecule $T_{cc\bar{s}}^+$ we defined here may exist.

We are mainly interested in the production of the $T_{cc\bar{s}}^+$ state in this work. In the $B_c^+$ decays, the production mechanism of $T_{cc\bar{s}}^+$ is rather similar with that of $T_{cc}^+$. Replacing the $D^{(*)+}$ in Figs.~\ref{feyn:D0} and \ref{feyn:Dstar0} with $D^{(*)+}_s$, the rescattering diagrams contributing to the $B_c^+\to T_{cc\bar{s}}^+\bar{D}^0$ and $B_c^+\to T_{cc\bar{s}}^+\bar{D}^{*0}$ are illustrated in Figs.~\ref{feyn:TccsD} and \ref{feyn:TccsDstar}, respectively.

In the rescattering diagrams, the effective Hamiltonian governing the weak process $B_c^+\to M_{c\bar{c}} {D}^{(*)+}_s $ reads
\begin{eqnarray}
H_{eff}&=&\frac{G_F}{\sqrt{2}}V_{cb}^* V_{cs}\big[c_1 (\bar{b}c)_{V-A}(\bar{c}s)_{V-A} \nonumber \\
&&+c_2 (\bar{c}c)_{V-A}(\bar{b}s)_{V-A} \big],
\end{eqnarray}
with $|V_{cs}|=0.975$~\cite{ParticleDataGroup:2022pth}. The decay constants in the factorized amplitudes are taken as $f_{D_s}=0.247$ GeV~\cite{ParticleDataGroup:2022pth} and $f_{D^*_s}=0.272$ GeV~\cite{Becirevic:1998ua}. Notice that $B_c^+\to M_{c\bar{c}} {D}^{(*)+}_s $ is a Cabibbo-favored process compared with $B_c^+\to M_{c\bar{c}} {D}^{(*)+} $. Therefore the branching ratio of $B_c^+\to T_{cc\bar{s}}^+\bar{D}^{(*)0}$ via the rescattering processes is also expected to be larger.

\begin{figure}[htb]
		\centering
		\includegraphics[width=0.8\linewidth]{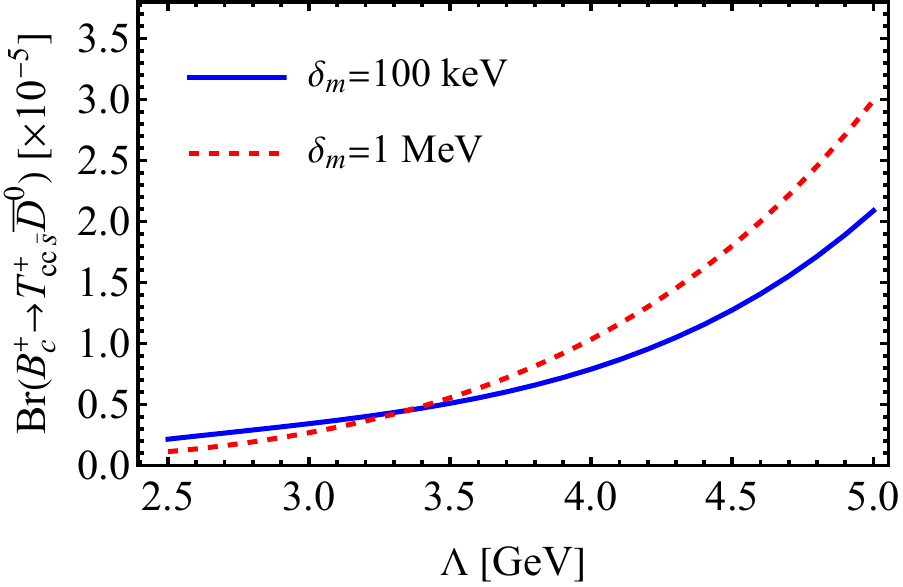}
		\caption{$\Lambda$-dependence of the branching ratio of $B_c^+\to T_{cc\bar{s}}^+ \bar{D}^0 $ via the rescattering processes in Fig.~\ref{feyn:TccsD}. }
		\label{fig:D0Tccs}
\end{figure}

\begin{figure}[htb]
		\centering
		\includegraphics[width=0.8\linewidth]{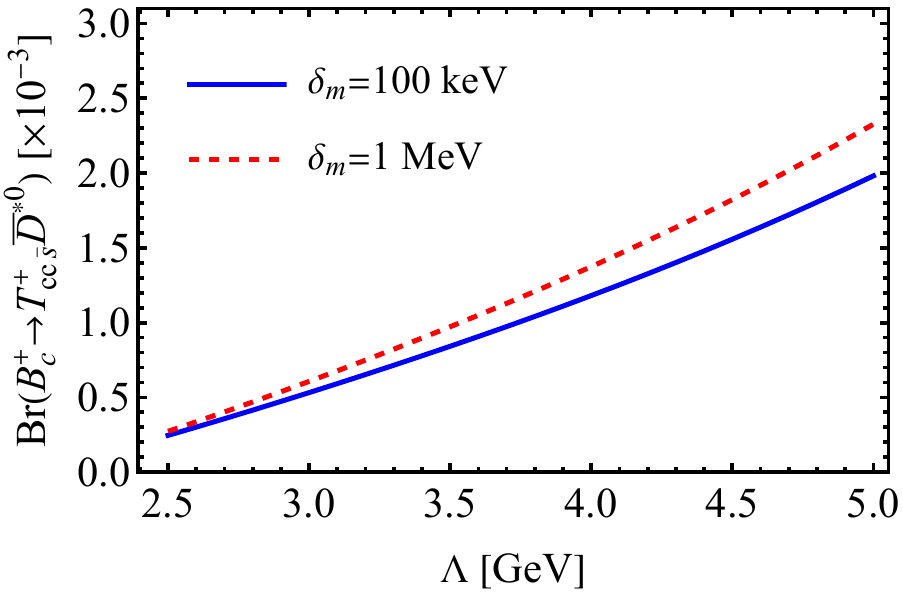}
		\caption{$\Lambda$-dependence of the branching ratio of $B_c^+\to T_{cc\bar{s}}^+ \bar{D}^{*0} $ via the rescattering processes in Fig.~\ref{feyn:TccsDstar}.}
		\label{fig:Dstar0Tccs}
\end{figure}

Following the rather similar procedures as that in Section \ref{sec:II}, we calculate the rescattering contributions and the numerical results are presented in Figs.~\ref{fig:D0Tccs} and \ref{fig:Dstar0Tccs}. We choose two typical values $\delta_m=100$ keV and $\delta_m=1$ MeV for a weakly bound hadronic molecule to estimate the couplings between $T_{cc\bar{s}}^+$ and its components. For the $B_c^+\to T_{cc\bar{s}}^+\bar{D}^{0}$ process, varying the $\Lambda$ from 2.5 to 5 GeV, the branching ratio increases from $\mathcal{O}(10^{-6})$ to $\mathcal{O}(10^{-5})$ for both $\delta_m=$100 keV and 1 MeV.
For the moderate cutoff $\Lambda$ around 3 GeV, $\mbox{Br}(B_c^+\to T_{cc\bar{s}}^+\bar{D}^{0})$ is of the order of $10^{-6}$. 
Compared with $B_c^+\to T_{cc\bar{s}}^+\bar{D}^{0}$, the $B_c^+\to T_{cc\bar{s}}^+\bar{D}^{*0}$ process has a larger branching ratio. The argument is similar as that in Section \ref{sec:II}. For both of the two $\delta_m$ values,   $\mbox{Br}(B_c^+\to T_{cc\bar{s}}^+\bar{D}^{*0})$ increases from $\mathcal{O}(10^{-4})$ to $\mathcal{O}(10^{-3})$ with $\Lambda$ increasing.  For the moderate cutoff $\Lambda$ around 3 GeV, $\mbox{Br}(B_c^+\to T_{cc\bar{s}}^+\bar{D}^{*0})$ is of the order of $10^{-4}$. This is a sizable branching ratio. If the $T_{cc\bar{s}}^+$ state truly exists, it is very likely that we can find it in the $B_c$ decays.

The two lines in Fig.~\ref{fig:D0Tccs} has a crossing point. This is because that the $T_{cc\bar{s}}^+ D^{*0}D_s^+$ coupling is sensitive to the $\delta_m$ when $\delta_m$ is smaller. Employing the similar coupling definitions as those in Eqs.~(\ref{eq:g1}) and (\ref{eq:g2}), The $T_{cc\bar{s}}^+ D^{*+}_s D^0$ coupling $g_1$ and  $T_{cc\bar{s}}^+ D^{*0}D_s^+$ coupling $g_2$ are determined to be $(g_1,g_2)\simeq(7.1\ \mbox{GeV},3.4\ \mbox{GeV})$ when  $\delta_m=100$ keV, and $(g_1,g_2)\simeq(7.8\ \mbox{GeV},6.0\ \mbox{GeV})$ when  $\delta_m=1$ MeV. As a result the interferences among rescattering diagrams for the two cases behave differently. The $\delta_m=100$ keV line is not simply scaled by the $\delta_m=1$ MeV line.


\section{Summary}
In this work, we study the production of the doubly charmed state $T_{cc}^+$ and its analog $T_{cc\bar{s}}^+$ in $B_c$ decays, which provide a good environment for the formation of the exotic meson containing double charm quarks.  The $T_{cc}^+$ or $T_{cc\bar{s}}^+$ is produced from a charmonium and a charmed meson reascattering via exchanging another charmed meson. The contributions from various rescatterings with different intermediate states are taken into account. The calculation of rescattering amplitudes is performed under the ansatz that $T_{cc}^+$ and $T_{cc\bar{s}}^+$ are weakly bound hadronic molecules. For the moderate cutoff energy, the branching ratios $\mbox{Br}(B_c^+\to T_{cc}^+ \bar{D}^{0})$ and $\mbox{Br}(B_c^+\to T_{cc}^+ \bar{D}^{*0})$ are estimated to be of the order of $10^{-7}$ and $10^{-5}$, respectively.
The $T_{cc\bar{s}}^+$ production in the $B_c^+$ decay is a Cabibbo-favored process. For the moderate cutoff energy $\mbox{Br}(B_c^+\to T_{cc\bar{s}}^+ \bar{D}^{0})$ and $\mbox{Br}(B_c^+\to T_{cc\bar{s}}^+ \bar{D}^{*0})$ are estimated to be of the order of $10^{-6}$ and $10^{-4}$, respectively.

We should also mention that even if $T_{cc}^+$, $T_{cc\bar{s}}^+$ and some other analogs are not hadronic molecules, the production mechanism of the doubly charm mesons proposed in this paper still works. Correspondingly, the couplings between the tetraquark states and open charm mesons need to be modified.

The predicted relatively sizable branching ratios suggest that in future experiments one may search for the $T_{cc}^+$ and its analogs in the $B_c$ decay processes discussed here. Besides, investigating the different production mechanisms of these exotic doubly charmed states is also crucial in revealing their underlying structures.

\begin{acknowledgments}
This work is supported by the National Natural Science Foundation of China (NSFC) under Grants No.~11975165, No.~12235018, No.~12175165 and No.~12075167. 
\end{acknowledgments}

\begin{appendix}	
\section{The vertex functions $\mathcal{A}(M_{c\bar{c}}\to \bar{D}^{(*)0} {D}^{(*)0})$ in the rescattering amplitudes}\label{sec:Appendix}
Here we give the vertex functions $\mathcal{A}(M_{c\bar{c}} (k_1)\to \bar{D}^{(*)0} (p_1) {D}^{(*)0} (k_3))$ corresponding to a charmonium coupling with a pair of open charm mesons:
\begin{eqnarray}
&&\mathcal{A}(J/\psi\to \bar{D}^{0}  {D}^{0}) \nonumber \\
&&= 2 g_\psi (k_3-p_1)\cdot \varepsilon_J m_D\sqrt{m_\psi}, \\
&&\mathcal{A}(J/\psi\to \bar{D}^{0}  {D}^{*0}) 
\nonumber \\
&&= -4i g_\psi \epsilon_{\mu\nu\alpha\beta} k_1^\mu p_1^\nu \varepsilon_{D^*}^{*\alpha} \varepsilon_J^\beta  \sqrt{\frac{m_D m_{D^*}}{m_\psi}}, \\
&&\mathcal{A}(J/\psi\to \bar{D}^{*0}  {D}^{0}) 
\nonumber \\
&&= -4i g_\psi \epsilon_{\mu\nu\alpha\beta} k_1^\mu p_1^\nu \varepsilon_{\bar{D}^*}^{*\alpha} \varepsilon_J^\beta  \sqrt{\frac{m_D m_{D^*}}{m_\psi}},\\
&&\mathcal{A}(J/\psi\to \bar{D}^{*0}  {D}^{*0}) \nonumber \\
&&= 2 g_\psi m_{D^*}\sqrt{m_\psi}[-(k_3-p_1)\cdot \varepsilon_J \varepsilon_{\bar{D}^*}^* \cdot \varepsilon_{D^*}^*  \\ \nonumber
&&+(k_3-p_1)\cdot \varepsilon_{D^*}^* \varepsilon_{\bar{D}^*}^* \cdot \varepsilon_J+(k_3-p_1)\cdot \varepsilon_{\bar{D}^*}^* \varepsilon_{D^*}^* \cdot \varepsilon_J] , \\
\nonumber
&&\mathcal{A}(\eta_c \to \bar{D}^{0}  {D}^{*0}) \nonumber \\
&&= 2 g_\psi (k_3-p_1)\cdot \varepsilon_{D^*}^* \sqrt{m_{\eta_c} m_D m_{D^*}}, \\
\nonumber
&&\mathcal{A}(\eta_c \to \bar{D}^{*0}  {D}^{0}) \nonumber \\
&&= -2 g_\psi (k_3-p_1)\cdot \varepsilon_{\bar{D}^*}^* \sqrt{m_{\eta_c} m_D m_{D^*}}, \\
\nonumber
&&\mathcal{A}(\eta_c \to \bar{D}^{*0}  {D}^{*0}) \nonumber \\
&&= -4i g_\psi \epsilon_{\mu\nu\alpha\beta} k_1^\mu p_1^\nu \varepsilon_{\bar{D}^*}^{*\alpha} \varepsilon_{D^*}^{*\beta} \frac{m_{D^*}}{\sqrt{m_{\eta_c}}} , \\
\nonumber
&&\mathcal{A}(h_c \to \bar{D}^{*0}  {D}^{*0}) \nonumber \\
&&= 2i g_\chi \epsilon_{\mu\nu\alpha\beta} k_1^\mu \varepsilon_{h_c}^\nu \varepsilon_{\bar{D}^*}^{*\alpha} \varepsilon_{D^*}^{*\beta} \frac{m_{D^*}}{\sqrt{m_{h_c}}}, \\
\nonumber
&&\mathcal{A}(h_c \to \bar{D}^{*0} D^0) \nonumber \\
&&= -2 g_\chi \varepsilon_{h_c} \cdot \varepsilon_{\bar{D}^*}^* \sqrt{m_{h_c} m_D m_{D^*}}, \\
\nonumber
&&\mathcal{A}(h_c \to \bar{D}^0 D^{*0}) \nonumber \\
&&= 2 g_\chi \varepsilon_{h_c} \cdot \varepsilon_{D^*}^* \sqrt{m_{h_c} m_D m_{D^*}}, \\
\nonumber
&&\mathcal{A}(\chi_{c0} \to \bar{D}^0 D^0) \nonumber \\
&&= -2 \sqrt{3} g_\chi m_D\sqrt{m_{\chi_{c0}}}, \\
\nonumber
&&\mathcal{A}(\chi_{c0} \to \bar{D}^{*0} D^{*0}) \nonumber \\
&&= \frac{2}{\sqrt{3}}  g_\chi \varepsilon_{\bar{D}^*}^* \cdot \varepsilon_{D^*}^* m_{D^*}\sqrt{m_{\chi_{c0}}}, \\
\nonumber
&&\mathcal{A}(\chi_{c1} \to \bar{D}^{*0} D^0) \nonumber \\
&&= 2\sqrt{2}i g_\chi \varepsilon_{\chi_{c1}} \cdot \varepsilon_{\bar{D}^*}^* \sqrt{m_{\chi_{c1}}m_Dm_{{D}^*}}, \\
\nonumber
&&\mathcal{A}(\chi_{c1} \to \bar{D}^0 D^{*0}) \nonumber \\
&&= 2\sqrt{2}i g_\chi \varepsilon_{\chi_{c1}} \cdot \varepsilon_{D^*}^* \sqrt{m_{\chi_{c1}}m_Dm_{D^*}}.
\end{eqnarray}

\end{appendix}

\end{document}